# Quantum computer of wire circuit architecture

S.A. Moiseev[1,2,3*], F.F. Gubaidullin[1,2] and S.N. Andrianov[1,3]

[1)] *Institute for Informatics of Tatarstan Academy of Sciences,*
*20 Mushtary, Kazan, 420012, Russia;*
[2)] *Kazan Physical-Technical Institute of the Russian Academy of Sciences*
*10/7 Sibirsky Trakt, Kazan, 420029, Russia;*
[3)] *Physical Department of Kazan State University,*
*Kremlevskaya 18, Kazan, 420008, Russia.*
*) samoi@yandex.ru

Quantum computer is an extremely urgent hardware for solution of complicated computational problems. There are various approaches to create quantum computer. First solid state quantum computer was built using artificial atoms – transmons (cooper pair boxes) in superconducting resonator. It was recently proposed to use an ensemble of atoms with long coherence time as a multi-qubit quantum memory for such kind of quantum computer where the cooper pair boxes are used for realization of single and two-qubit gates. However the operation of such computer is limited because of using a number of processing rigid cooper boxes working with fixed frequency at temperatures of superconducting material. Here, we propose a novel architecture of quantum computer based on a flexible wire circuit of many coupled quantum nodes containing controlled atomic (molecular) ensembles. We demonstrate wide opportunities of the proposed computer. Firstly, we reveal a perfect storage of external photon qubits to multi-mode quantum memory node and demonstrate a reversible exchange of the qubits between the quantum memory and any arbitrary nodes. We found optimal parameters of atoms in the circuit and self quantum modes for quantum processing. The predicted perfect storage has been observed experimentally for microwave radiation on the lithium phthalocyaninate molecule ensemble. Then also, for the first time we show a realization of the efficient basic two-qubit gate with direct coupling of two arbitrary nodes by using appropriate atomic frequency shifts in the circuit nodes. Proposed two-qubit gate runs with a speed drastically accelerated proportionally to the number of atoms in the node. The direct coupling and accelerated two-qubit gate can be realized for large number of the circuit nodes. Finally, we describe two and three-dimensional scalable architectures that pave the road to construction of universal multi-qubit quantum computer operating at room temperatures.

Construction of large quantum computer (QC) is a synergetic physical and engineering problem which imposes a number of critical requirements on physical and spatial organization of interconnections between the qubits of QC and with its near environment necessary for a quantum transmission and readout of quantum calculations results[1,2]. Quantum computing is based on delicate exploitation of number of various single- and two- qubit gates. Usually, single qubit gates are relatively easily fulfilled experimentally by using a well-known coherent control of single two-level atoms (atomic qubit) or molecular qubits in external resonant electromagnetic fields[5-7]. Also, the single photon qubit gates can be realized by the linear optics technique which provides simple procedures for rotation of the light phase and polarization with a control of light by mirrors and beamsplitters[8-9].

It is necessary to have a high enough coupling constant between qubits in order to realize sufficiently fast two-qubit gates. Few promising approaches have been proposed to increase a coupling constant of two qubit interactions. For example, using a single mode optical cavity provides the enhanced coupling constant of the interaction between an atomic qubit and a photon in the cavity[10,11]. This idea got an extensive use in modern proposals of quantum information processing. Very powerful method to increase the coupling constant with a photon is to use a Josephson qubits cooper characterized by superconducting current of mesoscopic magnitude[12] and by using the Josephson qubits (transmons) in superconducting resonators[13,14]. The two qubit transmon processor has been demonstrated recently[3] for successful implementation of the Grover search and Deutsch–Jozsa quantum algorithms.

Another promising tool of the coupling constant enhancement is an encoding of qubits on multi-atomic coherent states. Here, the coupling constant of N atoms with a photon can be enhanced by factor $\sqrt{N}$. Initially the multi-atomic coherent ensembles have been used for optical quantum memory (QM)[15-21], implementation of robust quantum communication over long lossy channels[22] and for single photon generation[23]. Recently collective ensembles of multilevel systems have been discussed for QM cooperated with quantum computing[24,25] and this idea has been developed to use the multi-qubit QM integrated in a hybrid superconducting QC for encoding of the qubits cooperated with transmon cooper pair box used for quantum processing[4]. In this paper, the single and two qubit gates can be performed by coherent control of transmon qubit in the external electromagnetic field and via the two-qubit iSWAP gate with the cavity photon state. Complete quantum computing for large number of qubits is realized by a transfer of any pair of the qubits on the transmon and photon qubits with their subsequent swapping and using single qubit gates. The hybrid approach had been proposed for quantum processing of more than 100 qubits stored in common atomic ensemble embedded in superconducting transmission line resonator.

**Wire circuit**

We propose a novel architecture of multi-qubit QC based on flexible wire circuit which couples many atomic ensembles situated in different spatial nodes. We demonstrate how the wire circuit architecture yields possibility to use available technologies for effective multi-qubit QM, faster quantum processing satisfying the di-Vincenzo criteria[26]. In particular, scalability is easily achievable for the proposed circuit based architecture. The simplest wire circuit scheme is shown in Fig.1. The circuit consists of three basic elements: an usual cupper *receiver* (R-) loop with diameter $D_1$, two-wire transmission (TWT-) line and second loop with highly reduced spatial diameter $D_2$ containing the resonant atoms (node). Two wires of the TWT-line are twisted in order to suppress an irradiation in external space. In our experiments, we used diameter $D_2 = 0.4$ mm which is more than 500 times smaller than the used wavelength $\lambda = 250$ mm of the radiation transmitted through the external waveguide. The R-loop provides an efficient reception of the radiation from the closely situated waveguide output loop as shown in Fig.1. At small length of TWT-line, the wire circuit works as a single mode resonator. We have been urged experimentally in a robust work of the wire resonator in its numerous spatial architectures. R-loop with inductivity L and capacity C tuned by varying length of line l (see Fig.1) determines a resonant frequency $\omega_o = 1/\sqrt{LC}$ of the circuit. TWT-line was adjusted to R-loop by choosing its spatial length as $l = n\pi c/(\omega_o\sqrt{\varepsilon_o \varepsilon}) = n\lambda_{TW}/2$, where $\lambda_{TW}$ is a wavelength of radiation in the TWT-line, c is the speed of light, $\varepsilon$ is a nondimensional permittivity of TWT-line volume, n is an integer. We note that a distance between two twisted wires in TWT-line was few orders smaller than the wavelength so the coupling between TWT-line and R-loop didn't change the resonant frequency $\omega_o$ while the line provided an effective transmission of the electromagnetic field between the loops. Since the RS-loop had a negligibly small spatial size the electromagnetic field evolved as a standing wave along the TWT-line with the same electrical current in the wires of nodes and R-loop. Thus the wire circuit works as a spatially distributed single mode resonator on the resonant frequency $\omega_o$ with Q-factor $Q = r^{-1}\sqrt{L/C}$, where r is a total losses resistance in the circuit (the losses were negligibly weak in our experiments) and the circuit provides an effective quantum electrodynamics of the atomic ensemble with electromagnetic field of the single resonator mode.

**Multi-mode quantum memory**

By following cavity mode formalism[27] we describe an efficient multi-qubit QM in our circuit. We use the Tavis-Cumming Hamiltonian[28,29] $\hat{H} = \hat{H}_o + \hat{H}_1$ for N atoms, field modes and their interactions generalized by taking into account inhomogeneous broadening of atomic frequencies and continuous spectral distribution of the field modes where $\hat{H}_o = \hbar\omega_o\{\sum_{j=1} S_z^j + \hat{a}^+\hat{a} + \sum_{n=1}^{2}\int \hat{b}_n^+(\omega)\hat{b}_n(\omega)d\omega\}$ are main energies of atoms ($S_z^j$ is a z-projection of the spin operator), energy of cavity field ($\hat{a}^+$ and $\hat{a}$ are arising and decreasing operators), energy of waveguide field (n=1) and energy of free space field (n=2) ($b_n^+$ and $b_n$ are arising and decreasing operators of the waveguide modes);

$$\hat{H}_1 = \hbar\sum_{j=1}\Delta_j S_z^j + \hbar\sum_{n=1}^{2}\int(\omega-\omega_o)\hat{b}_n^+(\omega)\hat{b}_n(\omega)d\omega$$
$$+ i\hbar g\sum_{j=1}[S_-^j\hat{a}^+ - S_+^j\hat{a}] + i\hbar\sum_{n=1}^{2}\int\kappa_n(\omega)[\hat{b}_n(\omega)\hat{a}^+ - \hat{b}_n^+(\omega)\hat{a}]d\omega. \quad (1)$$

The first two terms in (1) comprise a perturbation energies of atoms (where $\Delta_j$ is a frequency detuning of j-th atom) and the field modes; the third and fourth terms are the interaction energy of atoms with cavity mode ($S_+^j$ and $S_-^j$ are the transition spin operators, g is a coupling constant) and interaction energy of the cavity mode with the waveguide and free propagating modes characterized by coupling constants $\kappa_n(\omega)$.

We note that $[\hat{H}_o, \hat{H}_1] = 0$ and Hamiltonian $\hat{H}_o$ characterizes a total number of excitations in the atomic system and in the fields which is preserved during the quantum evolution where $\hat{H}_o$ gives a contribution only to the evolution of common phase of the wave function. $H_1$ determines a unitary operator $\hat{U}_1(t) = \exp\{-i\hat{H}_1 t/\hbar\}$ causing a coherent evolution of the atomic and field systems with dynamical exchange and entanglement of the excitations between them. In spite of huge complexity of the compound light-atoms system here we show that their quantum dynamics governed by $H_1$ in (1) can be perfectly reversed in time on our demand in a simple robust way.

We assume that initially all atoms ($j = 1,2,...,N$) stay on the ground state $|0\rangle = |0_1, 0_2, ..., 0_N\rangle$ and free mode fields are in the vacuum state and we launch a signal multi-mode weak field to the circuit through the waveguide at t=0 as shown in Fig.1. Total temporal duration of the signal field should be shorter than decoherence time of the atomic system ($\delta t \ll T_2$). By assuming that the spectral width $\delta\omega$ of the signal field is narrow in comparison with spectral window of waveguide transmission and inhomogeneous broadening of the resonant line ($\delta\omega \ll \Delta_{in}$) the consideration of the atoms and field evolution for $\delta t < t \ll T_2$ gives an efficiency $Q_{eff}$ of the signal field storage (a ratio of stored in atomic system energy to incoming energy of the signal field):

$$Q_{eff} = \frac{\gamma_1}{(\gamma_1 + \gamma_2)} \frac{4\Gamma/(\gamma_1 + \gamma_2)]}{|1 + \Gamma/(\gamma_1 + \gamma_2)]|^2}, \qquad (2)$$

plotted in Fig. 3, where $\gamma_i = \pi \kappa_i^2(\omega_o)$ determines coupling between the cavity mode with waveguide modes and with the free propagating modes, $\Gamma = N_{qm} g^2 / \Delta_{in}$ - coupling of cavity mode with atomic system, $N_{qm}$ is a number of atoms in the QM node.

As it is seen in Fig.2, the quantum efficiency $Q_{eff}$ reaches unity at $\Gamma/\gamma_1 = 1$ and $\gamma_2/\gamma_1 \ll 1$ that shows a possibility of perfect storage for multi-mode signal field at moderate atomic density. Note that $\gamma_1 = \Gamma$ is a condition of *optimal matching* between the waveguide modes and the atomic system in QM node. In this case all multi-mode signal fields incoming in the circuit transfer to the atomic system of the QM node. The efficient direct unconditional transfer of the multi-mode field is possible for inhomogenously broadened atomic (electron spin) transition where the effective quantum storage of multi-mode fields occurs for arbitrary temporal profile of the modes.

We examined experimentally the signal storage for the radiation field with carrier frequency $\nu = 1.2$ GHz. We varied parameter $\gamma_1$ by changing a spatial distance between R-loop and waveguide output loop in Fig.3 a) and found that all signal field energy was transferred to the electron spin systems of lithium phthalocyaninate (LiPc) molecule sample at the optimal matching value of $\gamma_1 = 3.768 \cdot 10^7$ where the reflected field was absent in the waveguide. At this condition we have observed a strong signal of the electron paramagnetic resonance depicted in Fig.4 b) from $\approx 4.51 \cdot 10^{13}$ LiPc molecules situated in one distant node.

In order to construct an efficient QM for the multi-mode fields we follow the original protocol of the photon echo QM proposed in 2001[19] and theoretically described[29] in most general way in the Schrödinger picture by exploiting symmetry properties of the light atoms Hamiltonian. Here, we exploit simplicity of this approach in description of the multi-mode QM in the proposed QC. The assumed atomic detunings $\Delta_j$ are caused by a presence of the magnetic field gradient. By assuming a perfect storage in accordance with above coupling matching condition we change a sign of the detunings $\Delta_j \rightarrow -\Delta_j$ at time moment t=t' by changing of the magnetic field polarity similar to recent experiments[30]. By using a substitution for the field operators $\hat{a} = -\hat{A}$ and $\hat{b}_n(\omega_o - \Delta\omega) = \hat{B}_n(\omega_o + \Delta\omega)$ (with similar relations for the Hermit conjugated operators) we get a new Hamiltonian $\hat{H}_1' = -\hat{H}_1$ which has an opposite sign with respect to initial Hamiltonian determining a reversed quantum evolution in accordance with a new unitary operator $\hat{U}_2[(t-t')] = \exp\{-i\hat{H}_1'(t-t')/\hbar\} = \exp\{i\hat{H}_1(t-t')/\hbar\}$. So the initial quantum state of the multi-mode signal field will be reproduced at $t = 2t'$ in the echo pulse with the irradiated field spectrum inverted relatively to the frequency $\omega_o$ in comparison with the original one. Thus we have shown for the first time that the multi-mode QM can be realized with 100% efficiency by using the optimal matching condition $\gamma_1 = \Gamma$ of the atoms in circuit with the waveguide modes.

**Quantum processing**

Let's consider a principle scheme of the QC operation for three node circuit depicted in Fig. 4. The circuit contains the QM node and two processing nodes. QM node is loaded in gradient magnetic field providing an inhomogeneous broadening of atomic frequencies $\Delta_{in} \gg \delta\omega_f$ with central atomic frequency coinciding with the circuit frequency $\omega_{qm} = \omega_o$. The second and third nodes have N atoms in each node with equal frequencies $\omega_{2,3}$ within each node tuned far away from the frequency $\omega_o$. We should provide a perfect transfer of arbitrary qubits between QM node and each other two nodes. Initially, the multi-qubit states encoded in the M temporally separated photon modes $E(t) = \sum_{m=1}^{M} E_m(t)$ with spectral width $\delta\omega_f$ are recorded in the QM-node by excitation from the external waveguide. When the storage procedure is completed we tune away the atomic frequency of the QM node from resonance with the circuit $\omega_{qm} \neq \omega_o$. In order to transfer one arbitrary k-th qubit state from the QM node to the second node we switch off the wire circuit coupling with the external waveguide ($\gamma_1 = 0$) and launch rephasing of the atomic coherence in QM node (by reversion of the atomic detunings $\Delta \to -\Delta$). The m-th qubit state will be rephased at t=2$t_m$. At time moment $t_k + t_{k-1}$ we equalize the frequencies of QM-node and 2-nd node with $\omega_o$. The quantum dynamics of atoms in QM and 2-nd nodes and circuit mode evolves to complete transfer of atomic excitation from the QM node to the second-node at t=2$t_k$ when the temporal shape of rephased single photon wave packet is $E_k(t) = E_o \exp\{-\Gamma |t - 2t_k|/2\} \sin S(t - 2t_k)/S$ ($S = \sqrt{Ng^2 - (\Gamma/2)^2}$, $t < 2t_k$, $E_o$ is an amplitude). These modes are the *self-modes* of the QC that provide the perfect reversible coupling of multi-mode QM with processing nodes. Similar temporal shape was proposed recently for single-mode QM[31]. After qubit transfer, we switch off the coupling of the 2-nd node with resonator by changing the frequency of atoms in the 2-nd node. The same procedure can be fulfilled for transfer of qubit from QM node to the 3$^{rd}$ node.

For realization of two-qubit gates we equalize the carrier frequencies of the 2-nd and 3-rd nodes with some detuning from the resonator mode frequency $\omega_{1k} - \omega_o = \Delta_{1k} = \omega_2 - \omega_o = \Delta_2 = \Delta$. This results in interaction of the atoms in the two nodes and between the nodes via the virtual processes of resonant circuit quanta determined by effective Hamiltonian[32,33] $\hat{H}_{eff} = \sum_{m=1}^{3} \hat{H}_{node}^{(m)} + \hat{H}_{int}$, where $\hat{H}_{node}^{(m)} = \hbar(|g|^2/\Delta) \sum_{i_m,j_m}^{N} S_{i_m}^+ S_{j_m}^-$ is a long-range spin-spin interaction in m-th node, $\hat{H}_{int} = \hbar(|g|^2/\Delta) \sum_{j_1,j_2=1}^{N} \left(S_{j_1}^+ S_{j_2}^- + S_{j_1}^- S_{j_2}^+\right)$ describes a spin-spin interaction between the nodes. The spin-spin interaction with atoms of QM-node is suppressed because of the absence of resonance.

Let's introduce four collective basis states in the two nodes: $|0\rangle_1|0\rangle_2$, $|1\rangle_1|0\rangle_2$, $|0\rangle_1|1\rangle_2$ and $|1\rangle_1|1\rangle_2$, where $|0\rangle_m = |0_1,0_2,...,0_{N_m}\rangle$ corresponds to ground state of the m-th node, $|1\rangle_m = (N_m)^{-1/2}\{|1_1,0_2,...,0_{N_m}\rangle + |0_1,1_2,...,0_{N_m}\rangle + ... + |0_1,0_2,...,1_{N_m}\rangle\}$ is a single atomic superposition excited in the 2-nd and 3-rd nodes (m=2,3) after transfer of a single photon state from QM node via the QC self-mode. It is important that the Hamiltonian $\hat{H}_{eff}$ has a matrix

representation in the basis separable from other states of the multi-atomic system ($N_2 = N_3 = N$)

$$\frac{\hbar |g|^2}{\Delta} \begin{pmatrix} 0 & 0 & 0 & 0 \\ 0 & N & N & 0 \\ 0 & N & N & 0 \\ 0 & 0 & 0 & 2N \end{pmatrix}. \qquad (3)$$

The structure of matrix (3) coincides with the matrix structure of two interacting two-level atoms. So the unitary evolution in the Hilbert space of four collective states will be the same but drastically accelerated N-times comparing to the case of two coupled two-level atoms. In particular the coherent oscillations between the states $|1\rangle_1|0\rangle_2$ and $|0\rangle_1|1\rangle_2$ will occur with coherent frequency $\omega_c = 2N|g|^2/\Delta$. By taking into account our experimental situation we can get $\Delta = 10\Delta_n$ that leads to $\omega_c = \gamma_1/10 = 3.768 \cdot 10^6$. It is known[32, 33] that the evolution of the two coupled two level atoms can lead to iSWAP and $\sqrt{\text{iSWAP}}$. In our case the gates are realized at shortened times: $1/\omega_c \cong 2.65 \cdot 10^{-7}$ sec. and $1/(2\omega_c) = 1.325 \cdot 10^{-7}$ sec. Thus using the described iSWAP realization we can achieve a fast transfer of a chosen qubit from the QM node to the second node. In the same way one can transfer arbitrary other qubit to the third node from the QM node. Then we can perform $\sqrt{\text{iSWAP}}$ gate which entangles the two qubits and provides a complete set of universal quantum gates together with single qubit operations. Single gates can be performed with arbitrary qubits by switching on the coupling $\gamma \to \gamma_1$ with a chosen node and transfer the atomic qubit to photonic qubit in waveguide where it can be rotated on arbitrary angle by usual means[9]. On demand we can return the qubit back to QM node by using well-developed fast methods of linear optics. Another possibility to organize single qubit gates is to transfer quantum information to the node with single resonant atom which state can be controlled by external classical field[11].

**Discussion**

The described principle 2D scheme of the quantum processing can be extended to large number of nodes as depicted in Fig.5. Moreover the 2D scheme is scalable to 3D architecture presented in Fig.6 due to its flexibility. We note an important possibility to make a parallel quantum processing with two different qubit pairs simultaneously by equalizing the carrier frequencies in each pair at different values between the pairs. In our experiments we have already realized a circuit with 5 spatially delocalized nodes organized along one straight TWT-line. We have observed strong EPR signals from each node that have demonstrated a real possibility to incorporate more than 100 coupled nodes in one QC of wire circuit architecture.

In this work we proposed a novel architecture of solid state QC based on wire circuit coupling multi-atomic systems in distinct quantum nodes. Experimentally we demonstrated basic physical properties of the architecture for perfect transfer of the light field to the circuit nodes and a possibility of a robust work with many distinct nodes. Theoretically we reveal perfect multi-mode QM for optimal experimentally attainable condition and we propose self-modes of QC and demonstrate a reversible transfer of an arbitrary qubit from the QM node to arbitrary other node. Also, we show a realization of fast iSWAP gate for arbitrary pair of two nodes which drastically accelerates basic quantum processes and opens a way for parallel operation of quantum gates. We demonstrated that the proposed QC architecture is a highly flexible for 2D and 3D wire-circuit configurations and that it is practically scalable for construction of many coupled distinct quantum circuits that opens a promising way for practical realization of multi-qubit QC with a number of coupled qubits limited only by the atomic coherence time. We anticipate that the proposed atomic QC can principally work at room temperatures for example on NV-centers in a pure diamond which looks now one of promising candidates of the qubit carriers up to $\sim 10^{-3}$-1 sec. timescale[34].

**Method Summary**

Experimental measurements have been performed on the original EPR spectrometer of L-band spectral range with resistive magnet of the Varian Company. Wire circuits were fabricated from a lacquered cupper wire with diameter 0.2 mm. Resonant frequency of circuit R-loop was tuned by using Rohde & Schwarz SMT03 (5kHz-3GHz) Sweep Generator. Radiation transfer from the waveguide to the circuit node has been controlled by detection of the minimum radiation reflection to the bridge of the spectrometer.

**Acknowledgment**

The authors thank the grant of the Russian Foundation for Basic Researches number 08-07-00449, Scientific School grant # 4531.2008.2 and Government contract of RosNauka # 02.740.11.01.03.

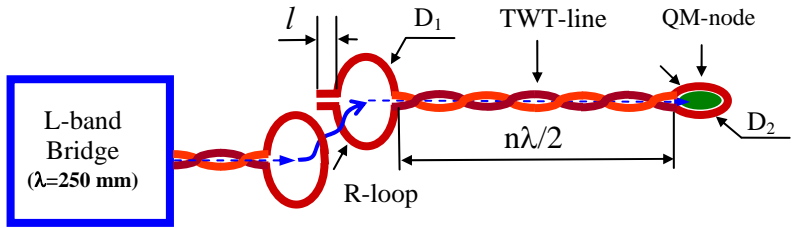

Figure 1. Experimental setup: L-band Bridge is a source of microwave radiation coupled with basic single node circuit. The circuit consists of capacity of the R-loop tuned experimentally by varying of length $l$, two wire transmission (TWT) line and small loop surrounding the resonant medium (green spot) - quantum memory (QM) node; $D_1$ and $D_2$ are the diameters ($D_2/D_1 \ll 1$). The radiation is transferred from the L-band Bridge to the - (QM) node along arrow.

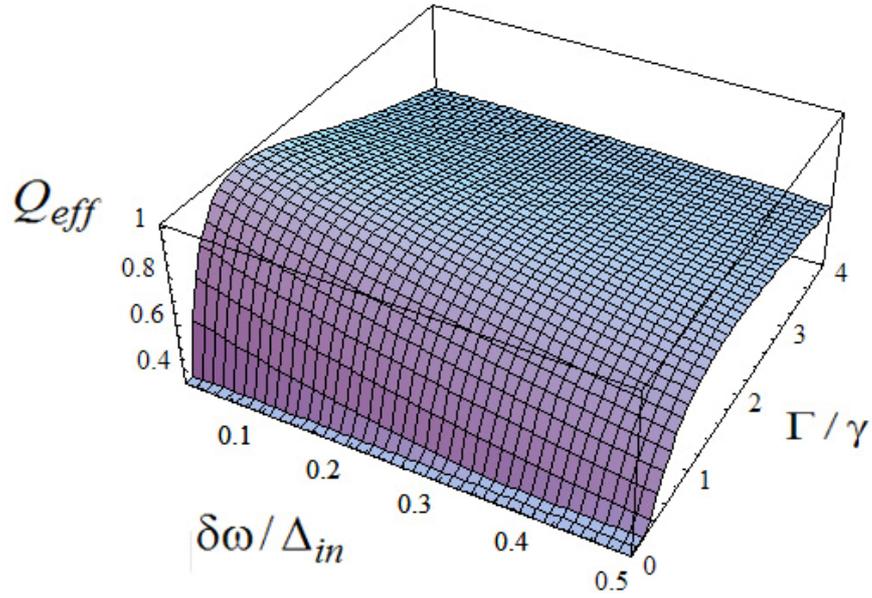

Figure 2. Transfer efficiency of the input light field to the QM node from external waveguide as a function of spectral width $\delta\omega/\Delta_{in}$ in units of inhomogenenous broadening and ratio of $\Gamma/\gamma$ for Lorenzian spectral shape of the input field for $\Delta_{in} = \gamma$. It is seen that $Q_{eff} > 0.9$ for $\delta\omega/\Delta_{in} < 0.2$ and the efficiency gets maximum at $\Gamma/\gamma = 1$.

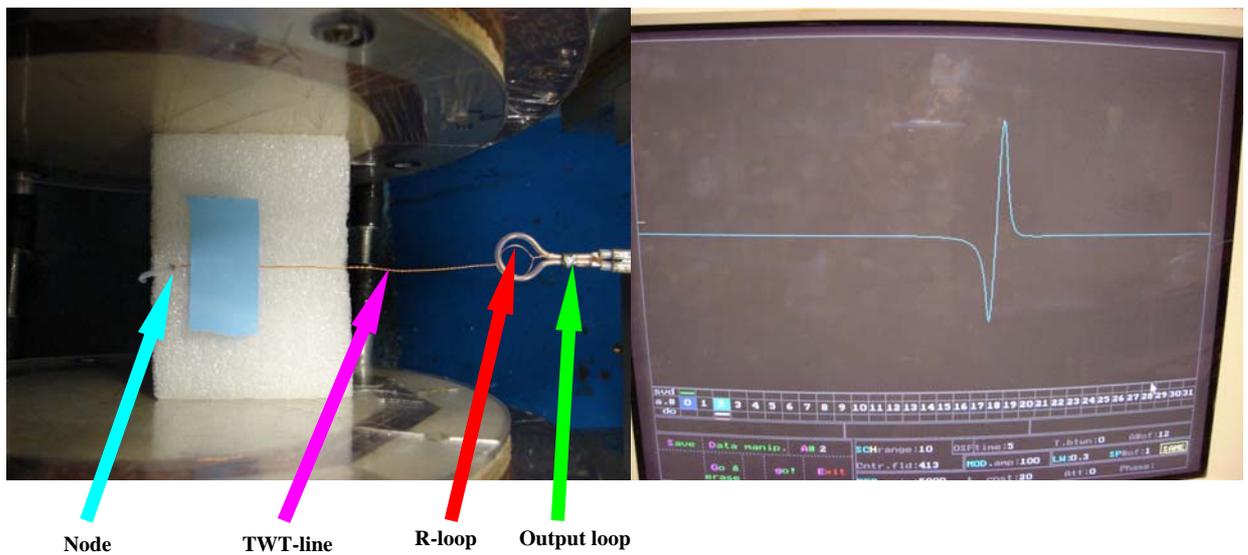

Fig.3. a) Efficient field transfer from R-loop to the spatially distant node was realized for sufficiently large TWT-line lengths (rose); R-loop, waveguide output loop and node are pointed by red, green and blue arrows; b) the strong EPR signal of LiPc spin system situated in the small size distant node has been observed at frequency $\nu = 1.2$ GHz with a spectral width $\Delta\nu = 88.2$ KHz.

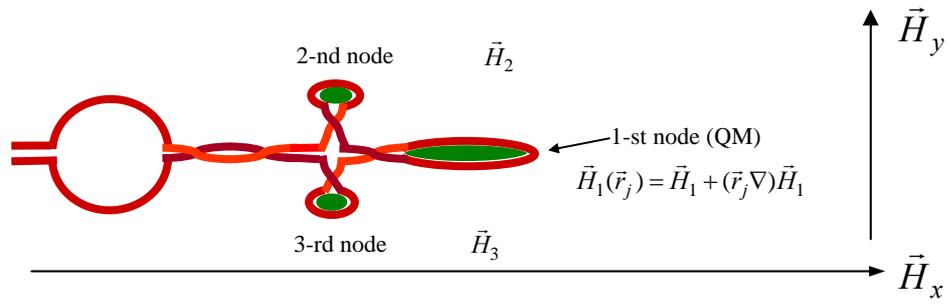

Figure.4. The 3-th node circuit. First node is a QM node loaded in gradient magnetic field; Second and third nodes are in constant different magnetic fields.

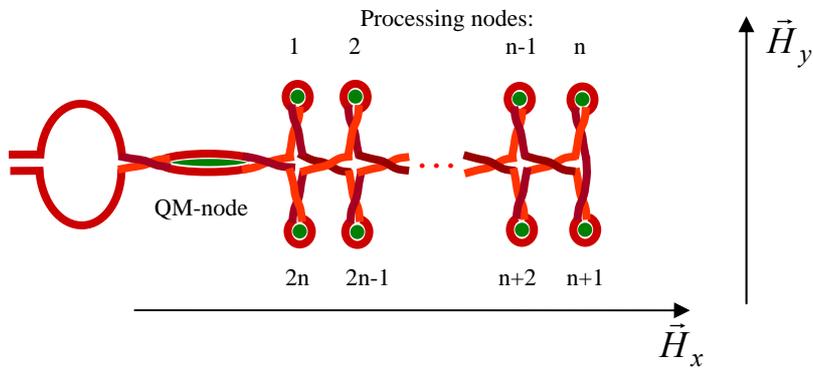

Figure.5. 2D-architecture of QC with one QM node and 2n processing nodes. Magnetic fields $\vec{H}_x$ and $\vec{H}_y$ are used to control atomic frequencies in the nodes; atoms in the QM node exist in the magnetic field gradient that leads to inhomogenenous broadening of the atomic frequencies. The magnetic field magnitude in the nodes is varied during the quantum processing for rephasing the atomic coherence and to control the resonance conditions in the circuit-node and node-node interactions.

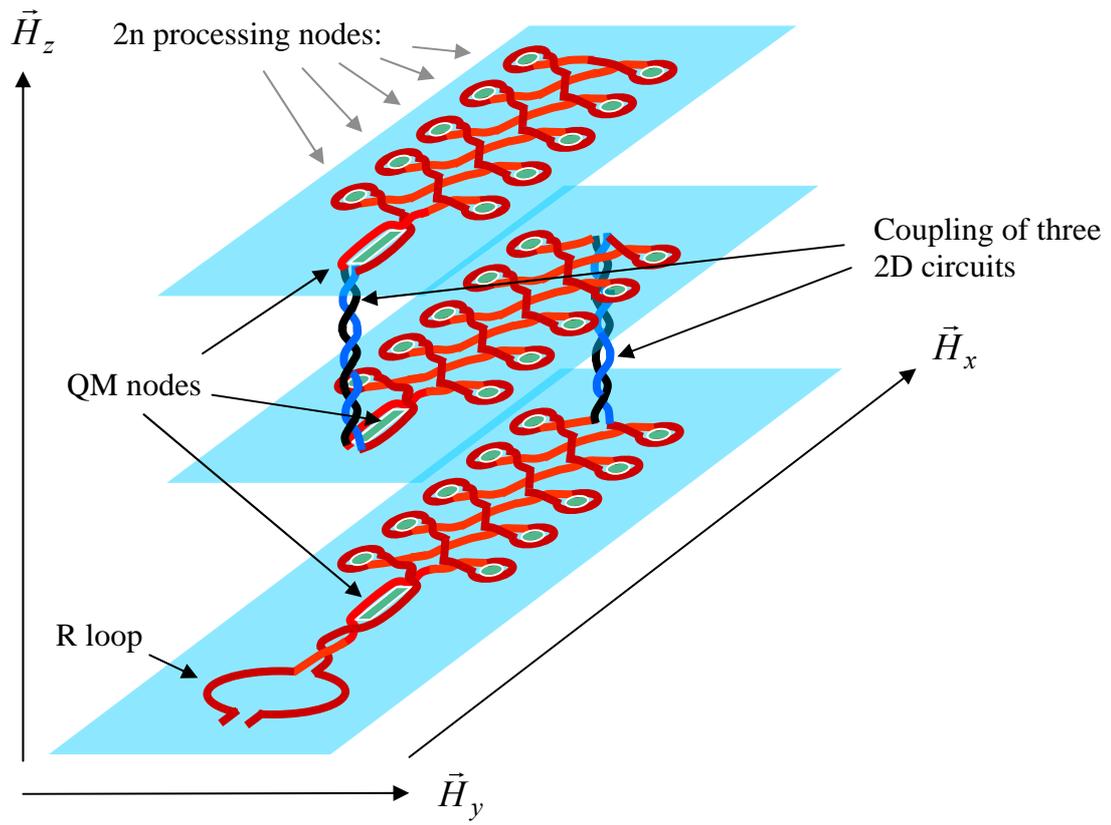

Figure 6. 3D scalable architecture of QC with 3 coupled 2D circuits with one receiver (R-) loop and three QM nodes and 3×2n processing nodes. Additional vertical TWT-lines are used for coupling of neighboring 2D circuits. Three components of the magnetic fields $\vec{H}_{x,y,z}$ can be used here for control of the atomic frequencies in all nodes.